1

# On the scaling law for cortical magnification factor.

Alexei A. Koulakov

*Cold Spring Harbor Laboratory, Cold Spring Harbor, NY 11724*

Primate visual system samples different parts of the world unevenly. The part of the visual scene corresponding to the eye center is represented densely, while away from the center the sampling becomes progressively sparser. Such distribution allows a more effective use of the limited transfer rate of the optic nerve, since animals can aim area centralis (AC) at the relevant position in the scene by performing saccadic eye movements. To locate a new saccade target the animal has to sample the corresponding region of the visual scene, away from AC. In this work we derive the sampling density away from AC, which optimizes the trajectory of saccadic eye movements. We obtain the scaling law for the sampling density as a function of eccentricity, which results from the evolutionary pressure to locate the target in the shortest time under the constraint of limited transfer rate of the optic nerve. In case of very small AC the visual scene is optimally represented by logarithmic conformal mapping, in which geometrically similar circular bands around AC are equally represented by the visual system. We also obtain corrections to the logarithmic scaling for the case of a finite AC and compare them to experimental findings.

The degree of sampling by the visual system is described by areal cortical magnification factor ($M$). It specifies the area in cortex (in mm$^2$), which represents a unit square of the image (in deg$^2$) (Daniel and Whitteridge, 1961). In primates the magnification factor is large in the AC and decays away from AC as a power law of eccentricity ($E$), measured in degrees

$$M = A/E^{2\alpha} \qquad (1)$$

Here $A$ is a constant, and $\alpha \approx 1$ is the scaling exponent determined from experiment (Van Essen et al., 1984). This paper is the first theoretical attempt to explain why the scaling exponent $\alpha$ has this value, and why in some cases it deviates from it.

An alternative measure of sampling is the density of retinal ganglion cells (RGC). It is natural therefore that magnification factor is related to RGC density. Away from AC the areal magnification factor $M$ and the density of ganglion cells $n_g$ are proportional, i.e.

$$n_g = B/E^{2\alpha} \qquad (2)$$

where $B$ is a constant, independent of eccentricity. The exponent $\alpha$ here is the same as in Eq. (1). This implies that each RGC projects to the same area in the primary visual cortex (V1).

Since each RGC projects to higher visual centers, the total number of RGC determines the thickness of optic nerve. Because a thick optic nerve impedes eye movements, the total number of ganglion cells is subject to a constraint (Meister, 1996). We assume that the total number of GC has reached its maximum possible value $N$, which does not substantially impair eye movements. Since this value is given by the integral of RGC density over the whole retina, it depends on two parameters in Eq. (2): $B$ and $\alpha$. Thus, this constraint does not allow calculating each of them individually: it only provides one condition on their combination (see Methods for more detail). To determine both of the two parameters unambiguously one has to find another condition.

Before formulating the other condition on $B$ and $\alpha$, we would like to provide an alternative motivation for the former anatomical constraint. Instead of fixing the total number of RGC we could fix the total area of cortex, represented by the integral of the magnification factor, given by Eq. (1), over the retinal space. Since Eqs. (1) and (2) are proportional the resulting anatomical constraints are equivalent.

What is the additional condition, which can fix both $\alpha$ and $B$ in Eq. (2) unambiguously? It is known that the oculomotor system in humans uses highly optimized strategy in the game of cricket, so that the cricket player's eye movement strategy contributes to his skill in the game (Land and McLeod, 2000). Eye movements also exhibit highly organized strategy during such everyday activities as tea making (Land et al., 1999) and driving (Land and Lee, 1994). Similarly, we suggest that the visual and oculomotor systems use mutually optimized strategy, which allows animals to detect new targets in the fastest way, whereby resulting in a fit organism, successful in the course of evolution.

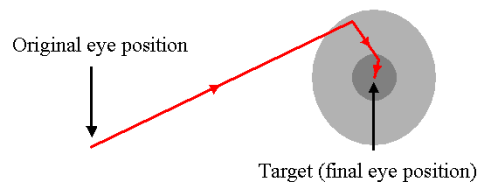

**Figure 1** Trajectory of AC in the visual space after an object suddenly appears at the target location. One primary and two corrective saccades are shown. The lighter gray disc schematically shows a region where the primary saccade lands in many trials due to a finite sampling density. The darker gray disk shows precision of the second saccade.

Consider the following basic task cooperatively performed by the visual and oculomotor systems (see Figure 1). Assume that the animal's AC is at certain position in the visual scene (original eye position). Then a new object (target) suddenly appears in the visual scene. We assume that the object is point-like for simplicity. The goal of the animal is to aim AC at the object for correct identification, performing saccadic eye movements. However, this cannot be done in one take, due to a sparse sampling on the periphery of visual scene. Indeed both visual acuity, represented by minimum angular resolution $\Delta$, and saccade precision depend on eccentricity as $\Delta = CE^\beta$, $\beta \approx 1$ (Weymouth, 1958; McKee and Nakayama, 1984). Thus, the saccade precision gets better when the target is closer to the AC, roughly linearly with eccentricity. After the first unsuccessful attempt to aim AC at the target the animal has much better chance with the second saccade, since the eccentricity of the target has decreased. The process repeats iteratively, until AC is at the target.

Comparing the sampling density, given by Eqs. (1) and (2) to the expression for angular resolution we notice that they are controlled by the same exponents α ≈ β ≈ 1. This implies that both minimum angular resolution and saccadic precision scale as distance between nearest ganglion cells: $\Delta \propto 1/\sqrt{n_g}$. Thus we can use the same exponent α to describe both the anatomical constraint derived from Eq. (1) and the iterative saccadic process shown in Figure 1. That we use the same scaling exponents for both saccadic precision (β) and magnification factor (α) implies that the same fraction of visual information flaw is dedicated to establishing the correct saccadic target locations for every target eccentricity. This assumption is based on the general uniformity of the visual system and is confirmed by the approximate equality between two scaling exponents, α and β.

We then minimize the total duration of the iterative process with respect to exponent α, subject to the anatomical constraint. In doing so we assume that saccades occur very fast and most of the time is consumed by a saccade preparation. Thus, we minimize the total number of saccades averaged over all possible target locations.

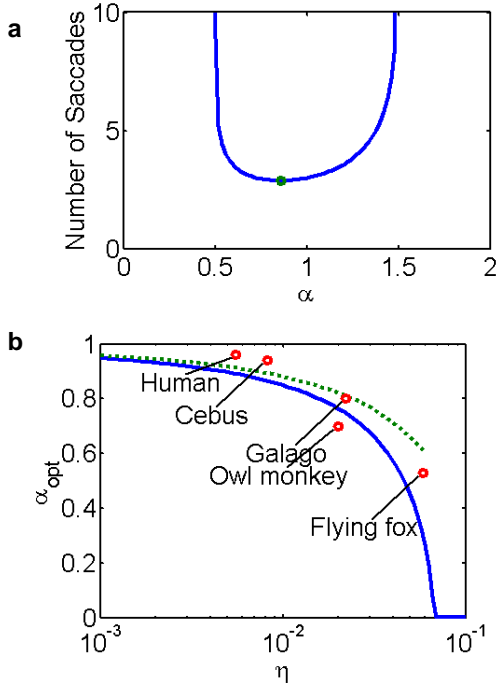

**Figure 2 a** The total number of saccades as a function of the scaling exponent. Parameters used are: N=600 and $\eta = 0.01$. The optimum is shown by the circle. The optimum value of $\alpha_{opt} \approx 0.9$ in which case the average number of saccades is about 3. **b** The optimum scaling exponent as a function of parameter $\eta$. The solid line is the result of our theory for N=600, and the dashed line is given by Eq. (3). The markers show experimental results for different species (see list in Methods)

This average number of saccades is shown in Figure 2a as a function of scaling exponent α. As seen from the figure the number of saccades diverges when α approaches 1.5. This is because in case of large α most of the ganglion cells are located near AC and the periphery of visual scene is undersampled. In this condition the objects on the periphery cannot be located reliably and the described iterative process includes a very large number of saccades. In the other extreme, when α ≈ 0.5, the periphery is well represented, but an object's location cannot be pinpointed exactly due to poor resolution near AC. The average saccade number has a minimum at α_opt ≈ 0.9, marked by a dot.

Our theory has two dimensionless parameters (see Methods). The first parameter describes the total number of "saccade pixels" in the visual scene, determined by the saccade precision. Since saccade precision is about 20% in humans, this parameter *N* is about 600, as estimated in Methods. This parameter describes the anatomical constraint and is controlled by the total number of ganglion cells. The second parameter is η = *r/R* << 1, where *r* ≈ 1° is the relative radius of AC and R ≈ 90°, is a measure of inhomogeneity introduced into the sampling density by the presence of AC. It therefore quantifies the finite size effects produced by AC. We derive expression for the value of α which optimizes the average saccade number for the most interesting case η << 1

$$\alpha_{opt} \approx 1 - \frac{3}{2}\left[\frac{\ln(g/\ln 1/\eta)}{\ln 1/\eta}\right]^2 \frac{1}{2+\ln(g/\ln 1/\eta)} \quad (3)$$

Here we introduced $g=N/2\pi \approx 100$ for brevity. The limit η → 0 corresponds a very small AC. When AC size (parameter η) becomes small, the second term in this expression becomes small too, and the optimum value of the exponent α approaches 1. This behavior is in agreement with the experimental observation of α ≈ 1. The second term in Eq. (3) describes the impact of finite (non zero) size of AC. In the simple theory presented here the correction is always negative. The comparison between theoretical and experimental results for some animals with finite AC is presented in Figure 2.

The value α = 1 corresponds to the logarithmic mapping of the visual scene. In this map the cortical position corresponding to eccentricity *E* is equal to ln *E*. The linear magnification factor resulting from the logarithmic sampling decays as $d\ln E/dE = 1/E$. The linear magnification factor for the second angular coordinate can be predicted from the assumption of conformal mapping, i.e. that small circles in the visual world are represented by circles in the brain and, therefore, the shapes of the objects are preserved by the brain map. The second linear magnification factor is therefore bound to be $1/E$ too, leading to the areal magnification factor *M* proportional to $1/E^2$. This reasoning results in α = 1 in Eq. (1). The conformal logarithmic mapping is optimal because the information learned about an object's location is equal between different saccades involved in targeting the object.

In conclusion, we suggest that distribution of sampling density existing in many primates is optimum for fast peripheral target location. The optimum is found under the constraint of limited transfer rate thorough the optic nerve.

**Methods**

*Finding the optimum scaling exponent*

We describe finding a new target by an iterative process, in which the length of the subsequent saccade is determined by the precision of the previous one: $l_{n+1} = \Delta(l_n)$. The expression for the dependence of saccadic precision on eccentricity is related to the ganglion cell density: $\Delta(l) = C/\sqrt{n_g(l)}$, where $C << 1$ determines the fraction of the visual bandwidth used to locate saccadic target. Using Eq. (2) we obtain: $\Delta(l) = l^\alpha C/B^{1/2} \equiv \tilde{C}l^\alpha$. Since $\alpha \approx 1$, $\tilde{C}$ has the meaning of relative saccade precision and is therefore about 0.2. The iterative sequence $l_{n+1} = \tilde{C} l_n^\alpha$ can be solved exactly to give

$$l_n = \tilde{C}^{\frac{1}{1-\alpha}} \left( l_0 / \tilde{C}^{\frac{1}{1-\alpha}} \right)^{\alpha^n},$$

where $n = 0, 1, 2, \ldots$ and $l_0$ is the length of the first saccade, which is approximately equal to the eccentricity of the target. The number of saccades needed to put fovea on the target is determined by assigning $l_n = r$, where $r$ is the radius of fovea. Averaging this expression over the target position $l_0$ between $r$ and $R$, which is the radius of retina in degrees, we obtain

$$n = \frac{1}{\ln \alpha} \ln \frac{\ln r/l_f}{\ln R/l_f}. \quad (M1)$$

Here

$$l_f = r \left[ \frac{N}{\pi} \frac{\alpha - 1}{1 - (r/R)^{2(\alpha-1)}} \right]^{\frac{1}{2(\alpha-1)}},$$

where

$$N = \frac{2\pi}{\tilde{C}^2} \ln \frac{R}{r} \quad (M2)$$

is the total number of saccadic "pixels", which is kept fixed during the optimization process. It represents a fraction of the total visual bandwidth, which can be used for locating targets, due to complexity of the task. The function $n = n(\alpha)$ is shown in Figure 2a. To derive Eq. (3) we expand Eq. (4) in a Taylor series around point $\alpha = 1$ to the second degree of parameter $\alpha - 1$. We therefore find the parabolic approximation of $n(\alpha)$ there. The minimum of the resulting parabola is easily found, which results in Eq. (3).

*Saccadic precision*

To obtain a realistic estimate for the saccadic precision one should measure saccadic errors in the natural conditions, in which animals compete for survival, such as presence of distracters, noisy background, weak target luminance etc. To the best of our knowledge such measurements have not been done. We estimate the lower bound for the saccadic errors from the measurements done in laboratory conditions. To mimic natural unpredictability of the target location one has to use experimental set up, in which both target eccentricity and direction are random and vary in a wide range. Such measurements have been done for human subjects (Deubel, 1985). The saccadic precision, which follows from this study, is about 20% of the target eccentricity (see Figure 3). This estimate is different from other estimates of about 10% precision (Becker and Fuch, 1969), due to difference in the task (random directions). We therefore adopt the following estimate to saccadic precision $\Delta(E) = 0.2 E$. The number of saccadic "pixels" which follows from this estimate [see Eq. ()] is $N \approx 700$.

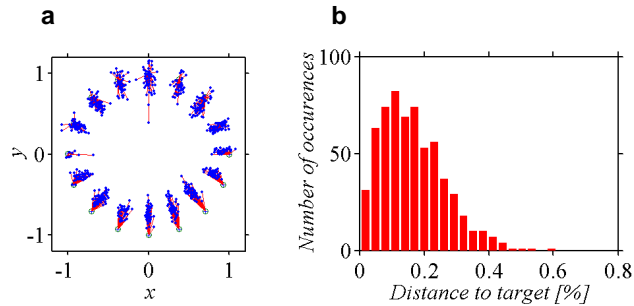

**Figure 3 a** Various positions at which saccades of a human subject landed (dots) relative to the target. The target is shown by a circle with a cross, red lines indicate saccadic errors. **b** The saccadic errors from **a** are shown in the histogram. The mean saccadic error, determining the saccadic precision is 17% of eccentricity. The data are from Deubel (1985).

*Comparison of the theory to primate data*

We use the following parameters for different primate species in Fig. 2b.

| Species | $r$ [deg] | $R$ [deg] | $\eta = \frac{r}{R}$ | $\alpha$ |
|---|---|---|---|---|
| *Galago*[1] | 2 | 90 | 0.022 | 0.8 |
| *Human*[2] | 1 | 90 | 0.011 | 0.96 |
| *Flying fox*[3] | 10 | 110 | 0.09 | 0.53 |
| *Owl monkey*[4] | 2 | 100 | 0.02 | 0.7 |
| *Cebus*[5] | 1 | 90 | 0.011 | 0.94 |

_________________________________________

_________________________________________

[1] Rosa et. al., 1997
[2] Weymouth, 1958; McKee and Nakayama, 1984
[3] Rosa et. al., 1993
[4] Silveira et. al., 1993
[5] Gattass et. al., 1987